\documentclass[aps,twocolumn,groupedaddress]{revtex4}

\usepackage{graphics}

\renewcommand{\(}{\begin{equation}}
\renewcommand{\)}{\end{equation}}

\begin{document}

\title{Quantum Coherence Beyond the Thermal Length}

\author{S. E. J. Shaw}

\affiliation{Department of Physics, Harvard University, Cambridge,
Massachusetts 02138, USA}

\author{R. Fleischmann}

\affiliation{Max-Planck Institut f\"ur Str\"omungsforschung,
Bunsenstra\ss e 10, D-37073 G\"ottingen, Germany}

\author{E. J. Heller}

\affiliation{Department of Physics and Department of Chemistry and
Chemical Biology, Harvard University, Cambridge, Massachusetts 02138,
USA}

\date{\today}

\begin{abstract}

Recent experiments have used scattering to map the flow of electrons
in a two-dimensional electron gas.  Among other things, the data from
these experiments show perseverance of regular interference fringes
beyond the kinematic thermal length.  These fringes are seen in full
quantum-mechanical simulations with thermal averaging, and within the
phase coherence length they can also be understood with a simple,
single-scattering model.  This effect provides a new way to gauge the
coherence length independent of thermal broadening.  Appealing to
higher-order scattering, we present a mechanism by which interference
fringes may survive even beyond the phase coherence length.

\end{abstract}


\maketitle

\section{Introduction}

In recent experiments, the flux of electrons through a quantum point
contact (QPC) and into the bulk of a two-dimensional electron gas
(2DEG) was probed with a charged atomic force microscope (AFM) tip
\cite{mat:science,mat:nature}.  The tip, capacitively coupled to the
2DEG, created a movable scatterer of the electrons propagating through
the system \cite{eriksson}.  The measurements taken were of the
conductance as a function of the position of the AFM tip above the
sample.  In the measurements, fringes spaced at half of the Fermi
wavelength and oriented transverse to the electron flux were seen.

Close to the QPC, we can understand these interference fringes as
arising from an open Fabry-P\'erot cavity between the tip and the QPC.
However, simple kinematic considerations suggest that these fringes
should die out at some distance from the QPC, as waves that differ in
energy drift out of phase with one another.  In the experimental
system, for example, waves differing in energy by $kT$ drift out of
phase by one radian over a round-trip distance of approximately
1300~nm.  The fringes seen experimentally, however, survive well
beyond this range.

\begin{figure}
\centerline{
	\raisebox{2truein}{A)}
	\resizebox{0.45\textwidth}{!}{\includegraphics{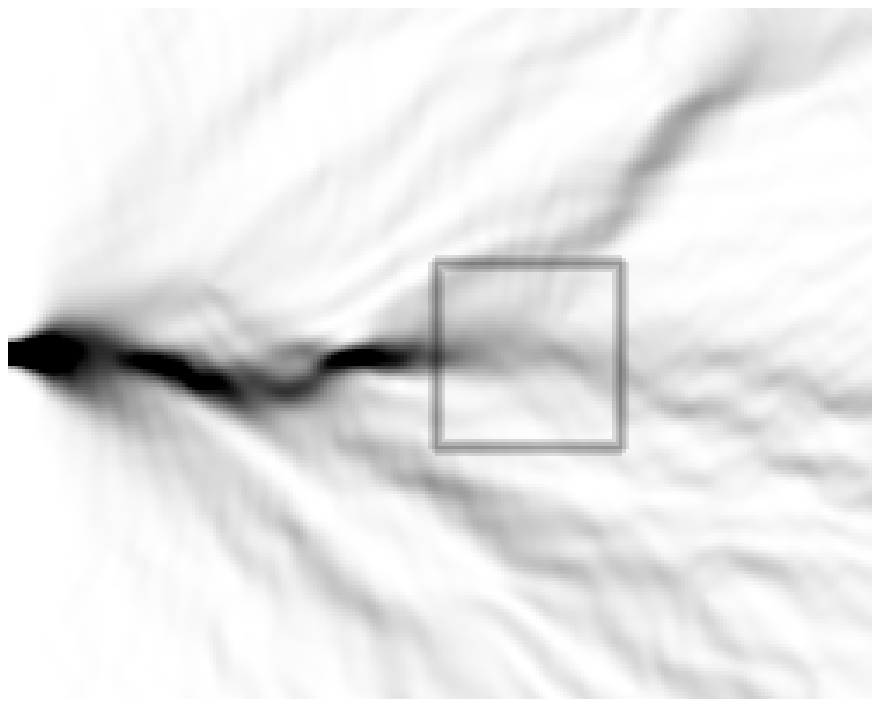}}
}
\vspace{4pt}
\centerline{
	\raisebox{1truein}{B)}
	\resizebox{0.2\textwidth}{!}{\includegraphics{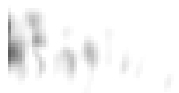}}
	\raisebox{1truein}{C)}
	\resizebox{0.2\textwidth}{!}{\includegraphics{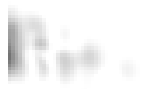}}
}
\caption{Here we see the survival of interference fringes beyond the
thermal length in a full quantum-mechanical simulation.  (A) shows the
quantum-mechanical flux through a system with a QPC and a disordered
background; for a discussion of the ``branched'' nature of the flux,
see~\cite{mat:nature}.  In (B) we show the results of a ``tip scan'',
i.e. conductance change as a function of AFM tip position, at a fixed
energy.  The scan is performed in the indicated area of (A).  In (C)
we show the thermally averaged tip scan data in the same region.  The
temperature for the average was set such that the scan region is two
thermal lengths from the QPC.}
\label{patchfig}
\end{figure}

The interference fringes beyond this thermal length are seen in full
quantum-mechanical simulations with thermal averaging, as shown in
Fig.~\ref{patchfig}, so we know that a complete theory reproduces
them.  This solution, however, doesn't tell us much about the
mechanism that allows the fringes to survive.  We present here a
simple model, appealing only to first-order scattering, that shows
that the fringes should survive up to the phase coherence length.
Another mechanism for qualitatively different fringes {\em beyond} the
coherence length is also presented; to date, experiments have not
probed this regime.

The potential seen by electrons in a 2DEG is fundamental to the model
presented here.  As reported in \cite{mat:nature}, we have considered
two contributions to this potential: impurities and donor atom density
fluctuations \cite{davies, grill}.  The contribution of the donor
atoms results in a random potential whose peaks lie well below the
energy of the electrons, and thus are not expected to backscatter
significantly.  The impurities, on the other hand, are occasionally
located close to the 2DEG and thus produce strong, localized
scattering centers.  It is these impurity scatterers that we will
consider in our models below.

\section{Single scattering}

Here we use a simple, first-order model to explain the fringes seen
beyond the thermal length.  The result requires phase coherent
transport at each energy, and therefore does not apply beyond the
phase coherence length.  This mechanism was previously mentioned in
\cite{mat:nature}, without the necessary details presented here, to
establish priority.

\subsection{Thermal averaging}

The thermal average is found by an integral over energy with the
derivative of the Fermi function as a weighting function.  In order to
simplify the mathematics of this model, we seek an approximation that
is an integral over wave vector with Gaussian weighting.  For the
ranges of parameters in this system, such an approximation can be made
to an acceptable degree of accuracy.  Typical values taken from the
experiments, for example, give us a temperature of 1.7~K and a Fermi
energy of 16~meV.

The thermal distribution of energies begins with the derivative of the
Fermi function at the known temperature $T$ and Fermi energy $E_F$.
We find that
\begin{eqnarray}
-f'(E)
& = &
\left[ 1 + e^{(E-E_F)/kT} \right]^{-2} \frac{1}{kT} e^{(E-E_F)/kT}
\nonumber
\\
& \approx &
\frac{1}{4kT} e^{-[(E-E_F)/(4kT \pi^{-1/2})]^2}
\label{weight1eq}
\\
& \approx &
\frac{1}{4kT} e^{-(k-k_0)^2 \ell_T^2}.
\label{weight2eq}
\end{eqnarray}
The standard deviation of the Gaussian in Eq.~(\ref{weight1eq}) was
chosen to match both the value of $-f'(E)$ at $E=E_F$ and its
approximate width; we can do both while preserving normalization.  In
Eq.~(\ref{weight2eq}) we have defined our kinematic thermal length
$\ell_T$ as
\( \ell_T = \hbar^2 k_0 \pi^{1/2} / 4 m kT. \label{lTeq} \)
%
This is half the distance (since we are interested in round-trips)
that it takes two waves separated in energy by the standard deviation
of the Gaussian in Eq.~(\ref{weight1eq}) to drift one radian out of
phase.  This then is the weighting function that we will use in
performing the thermal average.

Using Eq.~(\ref{weight2eq}) we will integrate over $k$ rather than $E$
in taking the thermal average.  Given the dispersion relation $E =
\hbar^2 k^2 / 2m$, we have $dE = (\hbar^2 k/m)\, dk$.  Appealing to
the physical values that will appear for $k_0$ and $\ell_T$, over the
range of the weighting function we can approximate this dispersion
relation as linear and take $dE = (\hbar^2 k_0/m)\, dk$.  Hence for a
signal $s(k, r)$ at fixed wave vector, we have the thermally averaged
signal $s(r)$ given by
\begin{eqnarray}
s(r) & = & \int \left( \frac{\hbar^2 k_0}{m} dk \right)
\frac{1}{4kT} e^{-(k-k_0)^2 \ell_T^2} s(k,r) \\
& = & \pi^{-1/2} \ell_T \int dk\ e^{-(k-k_0)^2 \ell_T^2} s(k,r).
\label{thermavgeq}
\end{eqnarray}

\subsection{The single scattering model}

This model is quite simple.  We take a random distribution of s-wave
scatterers in a plane at the points $\{\vec r_i\}$ and with scattering
lengths $\{a_i\}$.  We assume phase coherent transport over the
round-trip distances.  Furthermore, we make the following
approximations, which have no effect on the qualitative results and
little effect on the quantitative results: we use $r^{-1/2} e^{ikr}$
rather than Bessel functions for the two-dimensional s-waves, assume
for each scatterer a scattering amplitude proportional to the
scattering length, and assume a phase shift equal to the scattering
length times the wave number.  The quantity of interest is the
reduction of flux through the point contact as a result of the
scattering.

We take the QPC to have at least one channel open, and in the
single-scattering picture neglect backscattering from the point
contact.  Any conductance oscillations are then not due to
interference of the returning amplitude with the outgoing amplitude,
but rather result from interference of different ways of returning to
the QPC.  This is easily seen by considering an outgoing wave $\exp
[ikx]$ added to a backscattered wave $\epsilon \exp [-ikx +
\delta(r)]$ in the wire, where $\delta (r)$ is the phase shift due to
a backscattering obstruction at position $r$. It is readily seen that
the net flux is independent of $\delta$, meaning one source of
backscattering does not give fringes in conductance measurements.

Let the wave from the QPC be $r^{-1/2} e^{ikr}$, and the scattered
wave from a point scatterer, measured at the QPC, be $(c a_i/r_i)
e^{ik(2r_i+a_i)}$.  The constant of proportionality between the
scattering length and amplitude, $c$, depends on details of the
scattering potentials irrelevant to this model.  There are two factors
of $r_i^{-1/2}$, one for the falloff of the wave illuminating the
scatterer and one for the falloff of the scattered wave.  The phase
advances by the round-trip distance plus the phase shift.  Let the tip
be at a radius $r_t$ and have the scattering length $a_t$, giving a
similar return wave.  Finally, to simplify the notation, let $r_i'
\equiv r_i + a_i/2$.

The full return wave at a single energy is
\( \sum_i \frac{c a_i}{r_i} e^{2ikr_i'} + \frac{c a_t}{r_t}
e^{2ikr_t'}. \)
The absolute square of this wave, the interference of various return
paths, is the signal we require.  We concentrate on the cross terms,
which will give rise to the oscillations with $r_t$.  The cross terms
are
\( s(r_t,k) = 2 {\rm\ Re\ } \left[ \sum_i \frac{c^2 a_i a_t}{r_i r_t}
e^{2ik(r_i'-r_t')} \right]. \)

We thermally average this signal using Eq.~(\ref{thermavgeq}).
Averaging after the absolute square so that it is an incoherent
sum, we have
\begin{eqnarray}
s(r_t)
& = &
2 \pi^{-1/2} \ell_T {\rm\ Re\ } \left[ \int dk\ e^{-(k-k_0)^2
\ell_T^2} \right. \times
\nonumber
\\
& &
\left. \sum_i \frac{c^2 a_i a_t}{r_i r_t} e^{2ik(r_i'-r_t')} \right].
\end{eqnarray}
Changing the order of summation and integration, performing the
Gaussian integral, and taking the real part, we are left with
\( s(r_t) = 2 \sum_i \frac{c^2 a_i a_t}{r_i r_t} \cos[2k_0(r_i'-r_t')]
e^{-(r_i'-r_t')^2/\ell_T^2}.
\label{signal1eq} \)

What remains is a contribution to the signal from all scatterers that
are within a thermal length of being the same distance from the QPC as
is the tip.  In this expression, there is nothing special about the
condition $r_t > l_T$.  The thermal length still plays a role in that
it determines the width of the band around $r_t'$ that contributes to
the thermally averaged signal.  Note that the fringes predicted by
this model are at half the Fermi wavelength, as observed.
Furthermore, the fringes will be oriented perpendicular to the
direction of electron flow, also as observed.

\begin{figure}[h!]
\centerline{
	A) \raisebox{2em}{\resizebox{0.45\textwidth}{!}
	{\rotatebox{-90}{\includegraphics{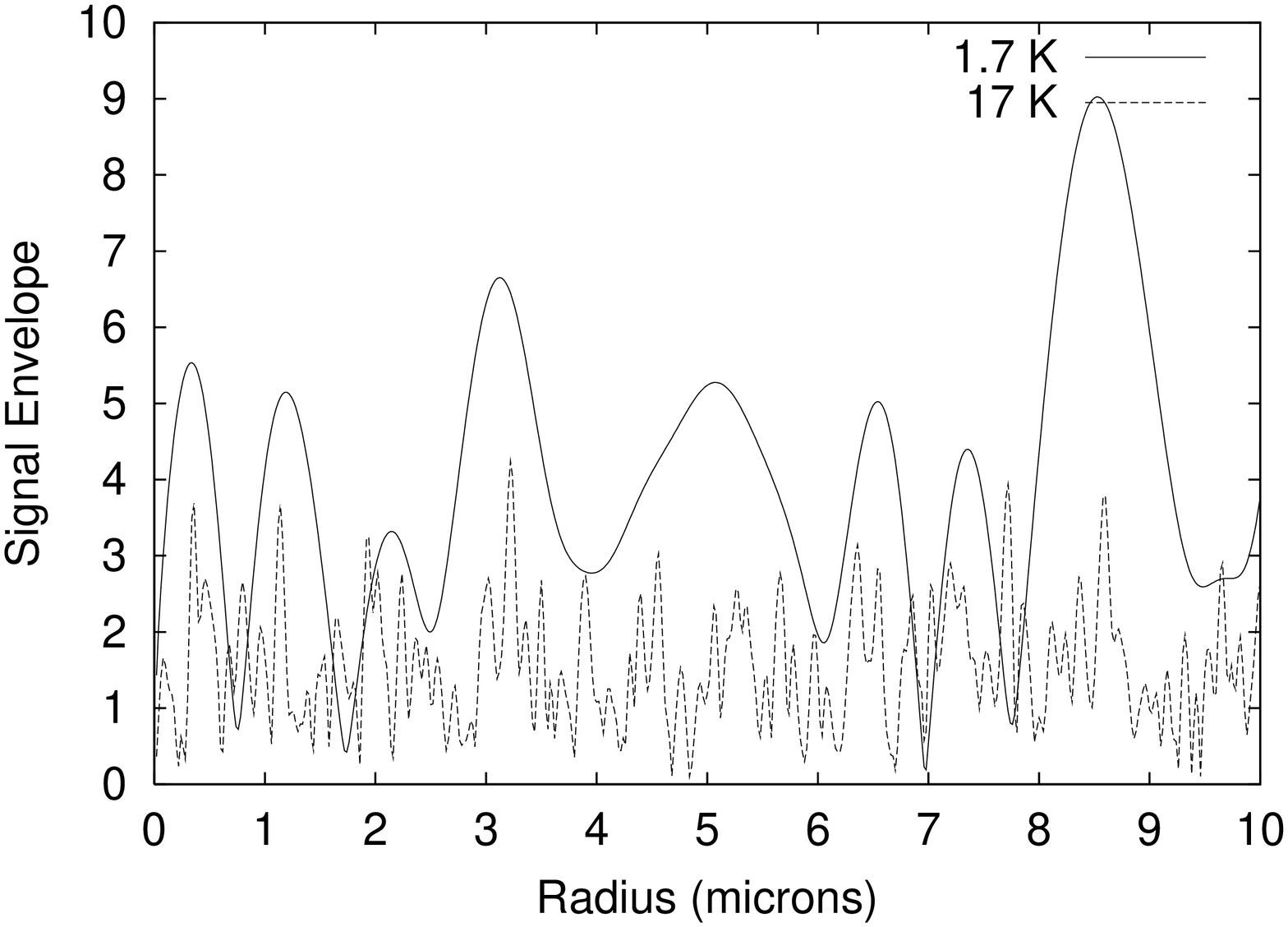}}}}
}
\centerline{
	B) \raisebox{2em}{\resizebox{0.45\textwidth}{!}
	{\rotatebox{-90}{\includegraphics{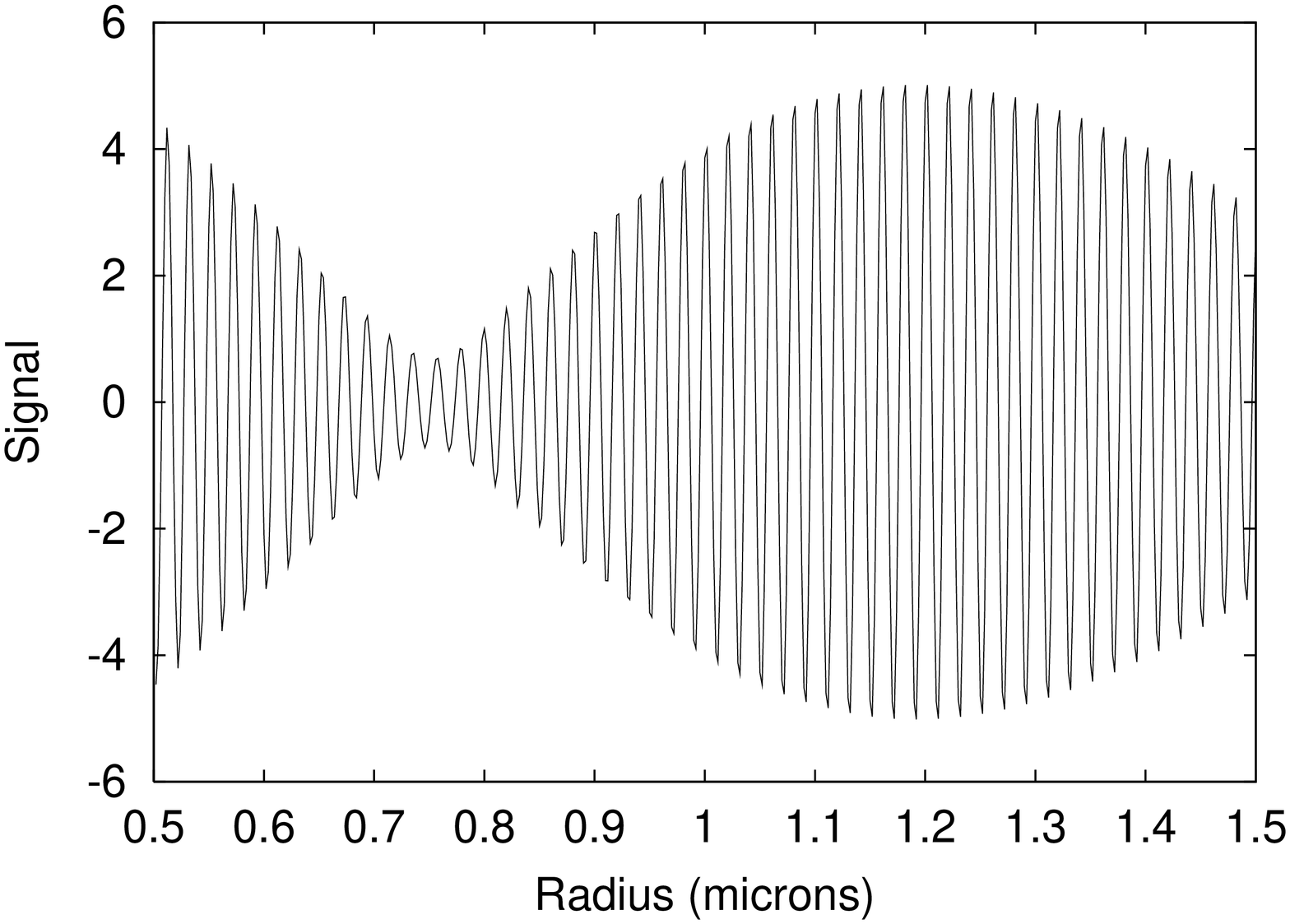}}}}
}
\caption{Several examples of the signal generated by
Eq.~(\ref{signal1eq}) with $\lambda_0 = 40 {\rm\,nm}$.  In all cases,
we have multiplied the data by $r^{3/2}$.  In (A) we show the
envelopes of two signals over a ten micron range; over this range, the
oscillations under the envelopes would not be discernible in the
figure.  The two signals show the same random distribution of
scatterers with a density of $40 {\rm\,\mu m^{-2}}$, and differ only
in their temperatures.  We note that the increase in temperature
results in a weaker signal with more rapid variation.  In (B) we show
the oscillatory signal under the envelope of the 1.7 K signal from
(A), looking over a one micron range similar to that probed in
experiments.}
\label{bornfig}
\end{figure}

In Fig.~\ref{bornfig} we show some examples of $s(r_t)$.  To make the
signal easier to observe, we divide out the overall $r^{-3/2}$
dependence of the signal strength.  Eq.~(\ref{signal1eq}) shows a
clear $r^{-2}$ dependence, but has an additional, not so obvious
factor of $r^{1/2}$.  This additional factor is related to the radial
dependence of the number of scatterers in our $\ell_T$-wide band and
the expectation value of a random sum of cosines.

There are other coherence effects that survive thermal averaging, such
as weak localization, where the interference of time-reversed paths
leads to an increased backscattering rate \cite{sheng}.  In contrast,
the single-scattering effect discussed here originates from the
interference of distinct paths of similar length, which can either
increase or suppress backscattering.

The derivation above assumed infinite coherence length; however, it is
clear that this mechanism cannot operate past this length, since it
depends on interference of coherent backscattered amplitude.  The
appearance (or disappearance) of these backscattering fringes can
become a new experimental measure of the coherence length.

\subsection{Single scattering continuum limit}

If we change the sum in Eq.~(\ref{signal1eq}) to an integral over the
plane and let $a_i = a \ \forall i$, we go from a system of discrete
scattering centers to a continuum approximation with constant density.
We have
\begin{eqnarray}
& & \int_0^\infty dr\,\pi r \, \frac{2 c^2 a a_t}{r
r_t} \cos[2k_0(r'-r_t')] e^{-(r'-r_t')^2/\ell_T^2} \nonumber \\
& = & {\rm Re\ } \left[ \int_0^\infty dr\, \frac{2 \pi c^2 a
a_t}{r_t} e^{2ik_0(r'-r_t')} e^{-(r'-r_t')^2/\ell_T^2} \right] \\
& = &
\pi^{3/2} \ell_T \frac{c^2 a a_t}{r_t} e^{-k_0^2 \ell_T^2} \{ 1 +
\nonumber
\\
& &
{\rm Re\ } [ {\rm\, erf} \left(i k_0 \ell_T +
(r_t'-a/2)/\ell_T \right) ] \}.
\end{eqnarray}

The error function is oscillatory, but bounded in magnitude.  The
dominant term for the magnitude of this signal is the exponential
$e^{-k_0^2 \ell_T^2}$.  For the values applicable in the physical
systems that we have considered, this is vanishingly small; in this
continuum limit, the signal disappears.

This implicit assumption of completely regular scatterer distribution
isn't particularly justified, however. We already know from multiple
scattering theory that such a regular distribution can appear as
simply a mean field, effectively ``raising the floor'' of the
potential over which the electrons propagate.

Even without this concern, this continuum limit calculation also tells
us nothing about how that limit is approached.  Simply taking the
single-scattering model in Eq.~(\ref{signal1eq}) and increasing the
density $d$ of scatterers (without any correlation in their placement)
reveals a $d^{1/2}$ dependence of the signal strength.  Clearly,
however, at some density the single-scattering model ceases to be a
useful one.  A detailed study of how the infinite-order scattering
calculations will cross over between domains is yet to be performed.
The predictions of this model do hold, however, for infinite order
scattering calculations with scatterer densities comparable to the
experimental system.  Preliminary results (Kalben et al.) indicate
that the fringes are qualitatively similar even as multiple scattering
becomes important.

\section{Multiple scattering}

Just as simple kinematic considerations would lead one astray about
the importance of the thermal length $\ell_T$, so can one overestimate
the importance of the phase coherence length $\ell_\phi$.  Though the
model described above depends on phase coherent transport, a
resonance model predicts fringes of a qualitatively different nature
at distances greater than $\ell_\phi$.

If the distance between the AFM tip and another scatterer is less than
$\ell_T$ and $\ell_\phi$, there can exist a resonance between the two.
This resonance can increase or decrease the conductance through the
full system despite incoherent transport over the distance back to the
QPC.  We are, in fact, quite accustomed to current carrying the
signatures of resonance over great distances of incoherent transport,
as this is the situation whenever we connect equipment to a 2DEG with
room-temperature wires.  This result is relatively easy to demonstrate
in one dimension, using the transfer matrix formalism with dephasing
built in.

The fringes generated by this resonance mechanism would not be
centered on the QPC, but rather on the fixed scatterer participating
in the resonance.  This is the primary qualitative difference between
this model and the single-scattering model proposed above.  In the
parameter ranges relevant to recent experiments (i.e., scatterer
densities and tip ranges) we expect the single scattering mechanism to
dominate.


\begin{thebibliography}{1}
\expandafter\ifx\csname bibnamefont\endcsname\relax
  \def\bibnamefont#1{#1}\fi
\expandafter\ifx\csname bibfnamefont\endcsname\relax
  \def\bibfnamefont#1{#1}\fi
\expandafter\ifx\csname url\endcsname\relax
  \def\url#1{\texttt{#1}}\fi
\expandafter\ifx\csname urlprefix\endcsname\relax\def\urlprefix{URL }\fi
\providecommand{\bibinfo}[2]{#2}
\providecommand{\eprint}[2][]{\url{#2}}

\bibitem{mat:science}
\bibinfo{author}{\bibfnamefont{M.~A.} \bibnamefont{Topinka}},
  \bibinfo{author}{\bibfnamefont{B.~J.} \bibnamefont{LeRoy}},
  \bibinfo{author}{\bibfnamefont{S.~E.~J.} \bibnamefont{Shaw}},
  \bibinfo{author}{\bibfnamefont{R.~M.} \bibnamefont{Westervelt}},
  \bibinfo{author}{\bibfnamefont{R.}~\bibnamefont{Fleischmann}},
  \bibinfo{author}{\bibfnamefont{E.~J.} \bibnamefont{Heller}},
  \bibinfo{author}{\bibfnamefont{K.~D.} \bibnamefont{Maranowski}},
  \bibnamefont{and} \bibinfo{author}{\bibfnamefont{A.~C.}
  \bibnamefont{Gossard}}, \bibinfo{journal}{Science}
  \textbf{\bibinfo{volume}{269}}, \bibinfo{pages}{2323} (\bibinfo{year}{2000}).

\bibitem{mat:nature}
\bibinfo{author}{\bibfnamefont{M.~A.} \bibnamefont{Topinka}},
  \bibinfo{author}{\bibfnamefont{B.~J.} \bibnamefont{LeRoy}},
  \bibinfo{author}{\bibfnamefont{R.~M.} \bibnamefont{Westervelt}},
  \bibinfo{author}{\bibfnamefont{S.~E.~J.} \bibnamefont{Shaw}},
  \bibinfo{author}{\bibfnamefont{R.}~\bibnamefont{Fleischmann}},
  \bibinfo{author}{\bibfnamefont{E.~J.} \bibnamefont{Heller}},
  \bibinfo{author}{\bibfnamefont{K.~D.} \bibnamefont{Maranowski}},
  \bibnamefont{and} \bibinfo{author}{\bibfnamefont{A.~C.}
  \bibnamefont{Gossard}}, \bibinfo{journal}{Nature}
  \textbf{\bibinfo{volume}{410}}(\bibinfo{number}{6825}), \bibinfo{pages}{183}
  (\bibinfo{year}{2001}).

\bibitem{eriksson}
\bibinfo{author}{\bibfnamefont{M.~A.} \bibnamefont{Eriksson}},
  \bibinfo{author}{\bibfnamefont{R.~G.} \bibnamefont{Beck}},
  \bibinfo{author}{\bibfnamefont{M.}~\bibnamefont{Topinka}},
  \bibinfo{author}{\bibfnamefont{J.~A.} \bibnamefont{Katine}},
  \bibinfo{author}{\bibfnamefont{R.~M.} \bibnamefont{Westervelt}},
  \bibinfo{author}{\bibfnamefont{K.~L.} \bibnamefont{Campman}},
  \bibnamefont{and} \bibinfo{author}{\bibfnamefont{A.~C.}
  \bibnamefont{Gossard}}, \bibinfo{journal}{Appl. Phys. Lett.}
  \textbf{\bibinfo{volume}{69}}, \bibinfo{pages}{671} (\bibinfo{year}{1996}).

\bibitem{davies}
\bibinfo{author}{\bibfnamefont{J.~H.} \bibnamefont{Davies}},
  \emph{\bibinfo{title}{The Physics of Low-Dimensional Semi-Conductors}}
  (\bibinfo{publisher}{Cambridge University Press}, \bibinfo{year}{1997}).

\bibitem{grill}
\bibinfo{author}{\bibfnamefont{R.}~\bibnamefont{Grill}} \bibnamefont{and}
  \bibinfo{author}{\bibfnamefont{G.~H.} \bibnamefont{D{\"o}hler}},
  \bibinfo{journal}{Physical Review B}
  \textbf{\bibinfo{volume}{59}}(\bibinfo{number}{16}), \bibinfo{pages}{10769}
  (\bibinfo{year}{1999}).

\bibitem{sheng}
\bibinfo{author}{\bibfnamefont{P.}~\bibnamefont{Sheng}},
  \emph{\bibinfo{title}{Introduction to Wave Scattering, Localization, and
  Mesoscopic Phenomena}} (\bibinfo{publisher}{Academic Press},
  \bibinfo{year}{1995}).

\end{thebibliography}

\end{document}